\documentclass[11pt, oneside]{article}   	
\usepackage{setspace}
\usepackage[margin=1.0in]{geometry}                		
\usepackage{graphicx}				
\usepackage{amssymb}
\usepackage{amsmath}
\usepackage{amsthm}
\usepackage{mathtools}

\usepackage{siunitx}
\usepackage{booktabs} 

\usepackage{times}

\usepackage[
	   	     backend=biber, 
		     style=authoryear
		     ]{biblatex} 	
\usepackage{xcolor}
\usepackage[colorlinks]{hyperref}
\usepackage[nameinlink, capitalize]{cleveref}

\usepackage{authblk}
\title{Biosignal Analysis with Matching-Pursuit Based Adaptive Chirplet Transform}
\author[1]{Jie Cui \thanks{Corresponding author, email: richard.cui@utoronto.ca}}
\author[2]{Dinghui Wang}
\affil[1]{Institute of Biomaterials and Biomedical Engineering, University of Toronto, Toronto, ON M5S 3G9, Canada. }
\affil[2]{Barrow Neurological Institute, St. Joseph's Hospital and Medical Center, Phoenix, AZ 85013, USA}
\date{}							

\allowdisplaybreaks

\begin{document}

\maketitle

\begin{abstract}
Chirping phenomena, in which the instantaneous frequencies of a signal change with time, are abundant in signals related to biological systems.  Biosignals are non-stationary in nature and the time-frequency analysis is a viable tool to analyze them.  It is well understood that Gaussian chirplet function is critical in describing chirp signals.  Despite the theory of adaptive chirplet transform (ACT) has been established for more than two decades and is well accepted in the community of signal processing, application of ACT to bio-/biomedical signal analysis is still quite limited, probably because that the power of ACT, as an emerging tool for biosignal analysis, has not yet been fully appreciated by the researchers in the field of biomedical engineering.  In this paper, we describe a novel ACT algorithm based on the ``coarse-refinement" scheme. Namely, the initial estimate of a chirplet is implemented with the matching-pursuit (MP) algorithm and subsequently it is refined using the expectation-maximization (EM) algorithm, which we coin as MPEM algorithm.  We emphasize the robustness enhancement of the algorithm in face of noise, which is important to biosignal analysis, as they are usually embedded in strong background noise.  We then demonstrate the capability of our algorithm by applying it to the analysis of representative biosignals, including visual evoked potentials (bioelectrical signals), audible heart sounds and bat ultrasonic echolocation signals (bioacoustic signals), and human speech.  The results show that the MPEM algorithm provides more compact representation of signals under investigation and clearer visualization of their time-frequency structures, indicating considerable promise of ACT in biosignal analysis.  The MATLAB\textsuperscript{\textregistered} code repository is hosted on GitHub\textsuperscript{\textregistered} for free download (https://github.com/jiecui/mpact).  
\end{abstract}
{\bf Keywords:} Chirplet, Biosignal processing, Matching pursuit, Time-frequency analysis, Sparse representation, Noise robustness.

\section{Introduction} \label{sec:intro}
Chirping activity can be encountered in many natural signals.  A chirp is a signal, in which the instantaneous frequency changes as a function of time.  Chirps can not only arise in such a variety of physical phenomena as radar signals \autocite{Mann1991, Wang2003}, mechanical vibrations \autocite{Guo2006}, ultrasonic echoes \autocite{Lu2005}, seismic waveforms \autocite{Boashash1986}, transionospheric signals \autocite{Qian1995, Doser1997}, and gravitational waves \autocite{Jenet2000, Candes2008}, but also, more relevant to biomedical engineering, abound in biological systems. One may find chirping phenominon, for example, in complex bird songs of different bird species \autocite{Glotin2016}, in powerful whale vocalization \autocite{Bahoura2008, Glotin2016}, in wolf choruses signals \autocite{Dugnol2008}, in human speech \autocite{Kepesi2006}, in neural responses correlated with auditory cortical processes \autocite{Mercado2000}, and in electromagnetic field related to brain activities measured as general electroencephalograph (EEG) \autocite{Kus2013, Sanei2007} and event related potentials (ERPs) \autocite{Cui2006a, Cui2008}. 

It is well understood that the chirp function, particularly, the Gaussian chirplet \autocite{Mann1995}, is one of the most important functions to characterize the signals with variable frequency.  However, its applications are relatively limited, mainly because of two major obstacles.  (1) One difficulty is that the chirplet functions do not generally constitute an orthogonal basis; as such, the decomposition of a signal into the basis functions is not unique, and the optimal approximation of a signal by linearly expanding chirplet basis is a $NP$-hard problem \autocite{Davis1997}.  (2) The other one is that biosignals are generally recorded under a condition of low signal-to-noise ratio (SNR). For example, the signal of visual evoked potentials (VEPs) are typically measured under the condition of SNR $ \approx -10 $ dB for a single trial and SNR $ \approx 0 $ dB for an average signal of 50 trials \autocite{Regan1989, Cui2006b}.  Therefore, chirplet decomposition of a low-SNR biosignal is in essence an estimation problem.  To overcome these limitations in the analysis of biological signals, the solutions suggested by the previous studies may be roughly classified into two categories: single chirplet estimation from segmented signals and multi-component chirplet extraction.  For instance, in a study to classify the calls of North Atlantic blue whales, Bahoura et al. \autocite{Bahoura2008, Bahoura2012} first employed a bandpass filter to suppress the influence of background noise on the main frequency band of whale call and then adopted the chirplet transform to approximate the call with a single chirplet atom only.  This approach avoided the problem of multi-component extraction, but whale vocalization cannot be completely characterized.  To further reduce computational cost, some studies fixed the rate of frequency changing (i.e. chirp-rate) of the lone chirplet (e.g. \cite{Shaik2015}).  A similar approach of extracting a single chirplet atom after partitioning the original signals was also proposed in the analysis of bioacoustics \autocite{Glotin2016} and visual evoked potentials \autocite{Cui2006c, Cui2008}.  In general, a longterm signal was partitioned into contiguous, non-overlapping and equal length segments using rectangular truncations.  A single chirplet was then estimated from each segment and hence the entire signal was approximated by a sequence of non-overlapping chirplets.  The advantages of this approach is its capability of representing the main time-frequency features of a signal with relatively low cost of computation, which is often crucial in the cases of longterm monitoring and data compression.   However, the representation with a single chirplet is not able to characterize the signals with complex time-frequency structures, such as EEG, which have multiple major components within an overlapping time intervals.  For these signals, a multi-component decomposition is necessary.  Typically, the estimation of chirplets from multi-component biosignals involves a ``coarse-refinement" scheme.  For example, in the studies of quantification of frequency-changing characters of sleep spindles in EEG \autocite{Schonwald2011, Carvalho2014}, the authors obtained the initial estimation of one chirplet (\emph{coarse} step) by adopting the matching pursuit (MP) algorithm with Gabor logon (or atom) dictionaries \autocite{Mallat1993}. In this first step it was assumed that the chirp-rate of the estimated spindle was zero.  Subsequently, in the second step (\emph{refinement} step) the time spread and the rate of linear frequency change of the spindle were re-estimated with a procedure named ``ridge pursuit" \autocite{Gribonval2001}, in which the time-frequency structure around the initial estimated Gabor logon was further explored at several testing points and then the time-spread and chirp-rate were estimated through a fast parabolic interpolation.  After the estimation of one chirplet spindle, the component was extracted from the signal and the next one was estimated from the residual signal, as is proposed in the standard procedure of MP algorithm \autocite{Mallat1993}.  It is worth noting that Yin et al \autocite{Yin2002} proposed a similar, two-step approach of multi-component chirplet extraction.  Similar to Gribonval's method, the initial parameters of a chirplet were estimated by fixing the chirp-rate to zero.  Unlike the former method, however, after a clever regroup of the equations of the inner product between the residual signal and the chirplet, the parameters were refined with a conventional curve fitting, leading to the reduction of computational cost.  Essentially, both methods acquire the Gabor logon (chirp-rate = 0) as the initial guess and then refine the chirp-rate and other parameters of the chirplet with linear operations.  However, this \textit{Gabor-to-chirplet} approach generally lacks robustness to strong noise, which is relevant to biosignal processing, especially when the centers of crossing chirplet components are in vicinity on the time-frequency plane \autocite{Lyu2015}.  

In this article, we show a new approach to the multi-component chirplet decomposition for the analysis of signals with biological nature.  It inherits the idea of ``coarse-refinement" scheme.  Unlike the previous works, however, the initial estimates are obtained with MP algorithm using Gaussian \emph{chirplet} dictionary \autocite{Bultan1999}, rather than Gabor logons.  Furthermore, the estimation of the chirplets are further refined using the expectation-maximization (EM) algorithm \autocite{Dempster1977, Feder1988, Mann1992}. We refer to this approach as the MPEM algorithm. Our results indicate that this approach is superior especially in the case when the analyzed signal is embedded in strong noise and the signal components cross each other on the time-frequency plan.  Although the theories of chirplet transform and overcomplete representation are well known in the community of signal processing, their applications in the field of bio-/biomedical engineering is relatively new.  We thus provide adequate background information and a detailed description of the implementation in the next section.  Subsequently, we demonstrate the merits of the proposed method by numerical simulation and several applications to real biological data, followed by the discussion.  Finally, we summarize our work in the last section. The code of the algorithm is freely available via GitHub\textsuperscript{\textregistered} (https://github.com/jiecui/mpact) under the GNU (GPLv3) public license \footnote{For the latest details of GPLv3 license please refer to http://www.gnu.org.}.  

\section{Matching-pursuit based adaptive chirplet transform}
The chirplet transform was formulated as a generalization of Gabor and wavelet transforms in the early 1990s \autocite{Mann1991, Baraniuk1993, Mann1995}. There is particular interest in developing Gaussian chirplet transform (GCT), because the basis function, Gaussian chirplet, is implemented as a modification to the original Gabor logon function \autocite{Gabor1946}, and thus it inherits some beneficial properties. Particularly, thanks to the Gaussian envelope, Gaussian chirplet has the highest joint time-frequency resolution and its Wigner-Ville distribution (WVD) is non-negative \autocite{Cohen1995}.  Moreover, the mathematical manipulation of GCT is usually tractable. Thus, the GCT plays a unique role in time-frequency analysis.

\subsection{From ``wavelet" to ``chirplet"}
The wavelet transform has been proposed to partially overcome the problem of the fixed time-frequency resolution with the short-time Fourier transform (STFT). Here we denote an arbitrary piece of sinusoid resulting from a windowing operation as a ``wavelet", with the only constraint being that the windowing function is a Gaussian function and hence each ``wavelet" is in fact a Gabor logon. A family of basis ``wavelet" functions can then be derived from a mother ``wavelet" by applying to it two operations: scaling (or time-spread) and time translation. A ``wavelet" has good time resolution but poor frequency resolution in the higher frequency bands, and vice versa for the lower frequencies. This is the reason why the wavelet transform is well-suited for analyzing signals with discontinuity or abrupt changes. However, this property also means that the wavelet transform does not provide precise estimates of the time-frequency structures for signal components that do not match the tradeoff characteristics of the wavelet signal. The wavelet is not efficient in representing chirp-like signals either.

\begin{figure}[h]
\begin{center}
    \includegraphics[scale=0.51]{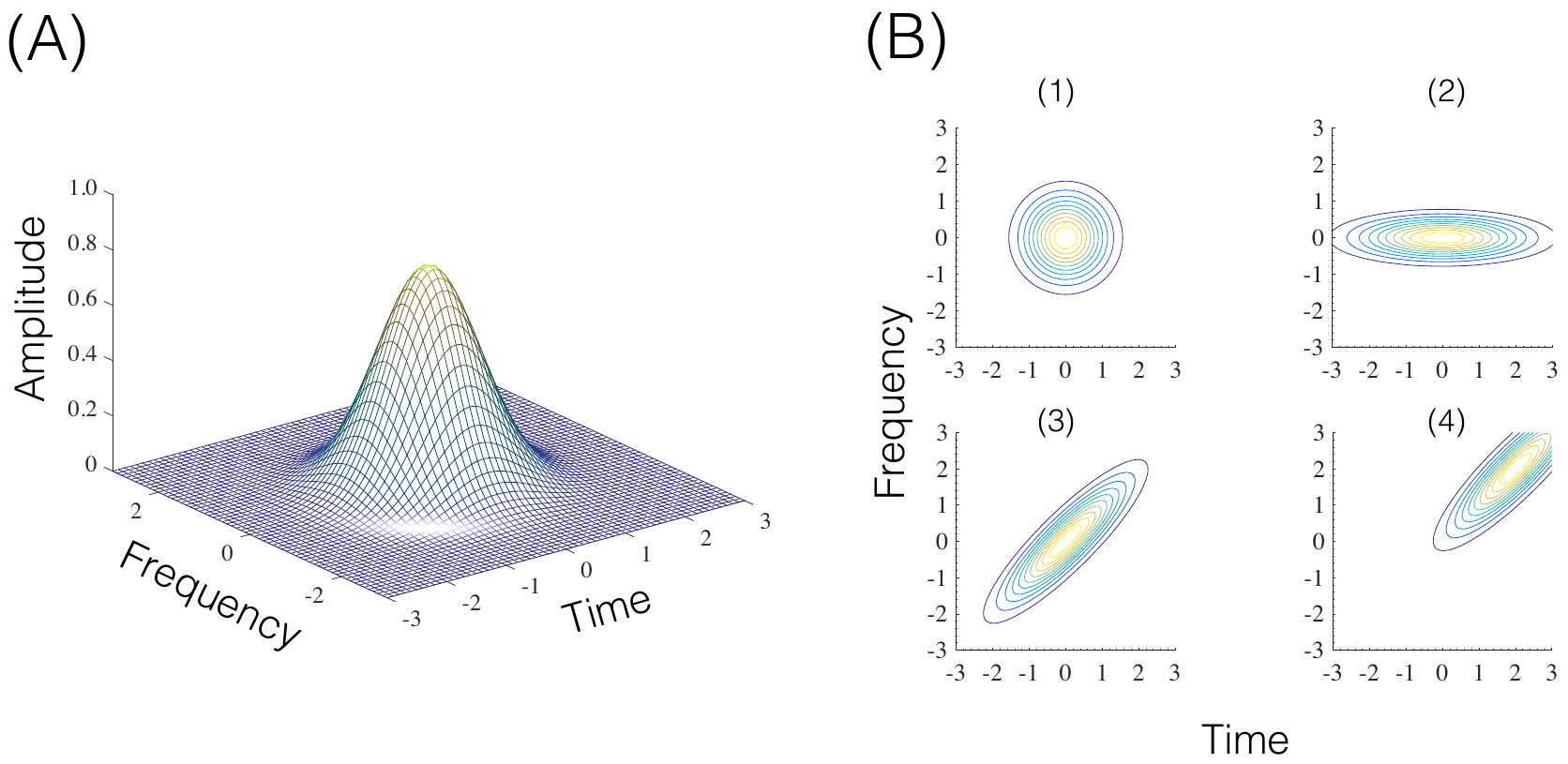}
\caption{\textbf{Construction of a Gaussian chirplet.} A chirplet may be constructed by applying the four mathematical operations to a unitary Gaussian function $ g(t) = \left( \pi \right)^{-1/4} \exp \left(  -t^2 / 2 \right) $. Panel \textbf{(A)} displays a 3-D visualization of the WVD of the unitary Gaussian function, while Panel \textbf{(B)} depicts the effect on (1) the unit Gaussian (represented as WVD contour) of the four operations, that is, (2) scaling $ \left( \pi \Delta_t^2 \right)^{-1/4} \exp \left[ -\left( t/\Delta_t \right)^2 /2 \right] $, (3) chirping $ \left( \pi \right)^{-1/4} \exp \left( -t^2 \right) \exp \left( j c t^2 \right) $, (4) time-shift $ \left( \pi \right)^{-1/4} \exp \left[ -(t - t_c)^2 \right] $, and frequency-shift $ \left( \pi \right)^{-1/4} \exp \left( -t^2 \right) \exp \left( j \omega_c t^2 \right) $, respectively.}
\label{fig:con_chirplet}
\end{center}
\end{figure}

In order to overcome these difficulties, the ``wavelet" is further modified to allow it to rotate in the time-frequency plane (the ``chirping" operation as introduced below). This is equivalent to windowing the chirp signal by using a Gaussian window and the resultant function is coined the ``chirplet". The chirplet can also be constructed from a unitary Gaussian function $ g(t) = \pi^{-1/4} \exp( -t^2/2) $ by applying four mathematical operations to it (\cref{fig:con_chirplet}, see \cref{eq:gaussain_chirplet} for the notation), i.e., 
\[
\begin{array}{ll}
  \text{(1) scaling} & (\pi \Delta_t^2)^{-1/4} \exp(-t^2/2\Delta_t^2),  \\
  \text{(2) chirping} & \pi^{-1/4} \exp( -t^2/2) \exp(jct^2), \\
  \text{(3) time-shift} & \pi^{-1/4} \exp[ -(t-t_c)^2/2], \\
  \text{(4) frequency-shift} & \pi^{-1/4} \exp( -t^2/2) \exp(j \omega_c t).
\end{array}
\]

A sequential application of these operations leads to a family of wave packets with four adjustable parameters called Gaussian chirplets
\begin{equation} \label{eq:gaussain_chirplet}
    g_{t_c, \omega_c, c, \Delta_t} (t) = \frac{1}{\sqrt{\sqrt{\pi} \Delta_t}} \exp\left\{ -\frac{1}{2} \left( \frac{t - t_c}{\Delta_t} \right)^2 + j\left[ c \left( t - t_c \right) + \omega_c \right] \left( t - t_c \right) \right\},
\end{equation}
where $ j = \sqrt{-1} $, $ t_c $ is the time center, $ \omega_c $ the frequency center, $ \Delta_t > 0 $ the effective time spread, and $ c $ the chirp-rate that characterizes the ``quickness" of frequency changes.  The effects of the four operations on the WVD of a chirplet are shown in \cref{fig:con_chirplet}.  It can be seen that the chirplet is simply a natural extension to ``wavelet" by applying an additional chirping or rotational operation. Indeed, both the Gabor logon and ``wavelet" are just special cases of chirplet, i.e., the case where the chirp-rate is zero.

\subsection{Gaussian chirplet transform and adaptive analysis} \label{sec:gct}
The Gaussian chirplet transform (GCT) of a signal is defined as the inner product between the signal
$ f(t) $ and the Gaussian chirplet $ g_{t_c, \omega_c, c, \Delta_t} (t) $ defined in \cref{eq:gaussain_chirplet}
\begin{equation} \label{eq:gct}
    a_I = \left< f, \, g_I \right> = \int_{-\infty}^{+\infty} f(t) g_I^*(t) \, dt, 
\end{equation}
where $ I = (t_c, \omega_c, c, \Delta_t) \in \mathbb{R}^3 \times \mathbb{R}^+ $ denotes the continuous index set of the chirplet parameters and `*' the complex conjugate operation.  The coefficient $ a_I $ is the projection of the signal $ f(t) $ onto a time-frequency region specified by the chirplet $ g_I $.  The absolute value of the coefficient is the amplitude of the projection.  An arbitrary signal can then be represented as a linear combination of Gaussian chirplets
\begin{equation}
    f(t) = \sum_{n = 1}^{P} a_{I_n} g_{I_n}(t) + R^P f(t) = f_P(t) + R^P f(t),
\end{equation}
where $ P $ is the number of chirplets, $ I_n $ is the parameter set of the $n$th chirplet, $  f_P(t) $ is defined as the $P$th-order approximation of the signal and $ R^P f(t) $ denotes the residue.  Notice that the coefficient $ a_{I_n} $ is complex and hence the decomposition information at each iteration $ n $ is described by six real parameters, i.e., two from $ a_{I_n} $ and the other four from $ I_n $.  The calculation of $ a_{I_n} $ involves selecting $ g_{I_n} $ from a predefined set of chirplets known as \emph{dictionary}. The approach is then to find the optimal subset of $ P $ chirplets from the dictionary so as to minimize the difference $ \| f - f_p \| $. Unfortunately, the optimal solution of $ a_{I_n} $ and $ I_n $ is an $ NP $-hard problem  \autocite{Davis1997}, i.e., there exists no known polynomial-time algorithm to solve this problem.  In practice, some suboptimal techniques have been developed (e.g., \cite{Qian1994, Bultan1999, Gribonval2001, Yin2002}) and we will describe one such approach.  The essence is to approximate the signalÕs energy in the time-frequency space using straight lines with arbitrary slopes (\cite{Cui2006b}, p.40). 

The general signal of interest here may be characterized by the following model, which assumes the signal is a sum of $ P $ chirplets in complex, white, Gaussian noise:
\begin{equation} \label{eq:sig_model}
f(t) = \sum_{n = 1}^{P} a_{I_n} g_{I_n}(t) + w(t),
\end{equation}
where $ w(t) $ denotes the additive noise.  We assume that any real signal of interest has been converted to a complex ({\it analytic}) signal, whose real part is the original signal and the imaginary part its Hilbert transform \autocite{Cohen1995, Flandrin1999}, so as to simply the theory. To decompose a given signal, two procedures are involved at each iterative step: (1) initial coarse estimates obtained using a \emph{chirplet} matching pursuit (MP) algorithm and (2) progressive refinement of the estimates with the estimation-maximization (EM) procedure  (\cref{tb:mpem}).  The initial stage of the algorithm includes the construction of a chirplet ``dictionary" and the initialization of the residue, $ R^0f = f $. A dictionary is a repertoire of chirplet basis functions selected to cover efficiently the entire time-frequency plane.  We follow the method proposed by Bultan \autocite{Bultan1999} and summarize the discretization of the parameters in \cref{tb:cons_dic}. 
\begin{table}[t]
    \caption{Discretization of chirplet parameters in dictionary construction}
    \begin{center}
        \begin{tabular}{ccl}
        
        \toprule[1.5pt]
        Symbol & Value & Description \\ 
        \midrule
        $ N $ & & Signal size (number of samples) \\
        $ i_0 $ & 1 (default) & The first level to chirp/rotate logons \\
        $ a $ & 2 (default) & Radix of scales \\
        $ \gamma $ & $ \gamma \in [0, N), \gamma \in \mathbb{Z} $ & Signal range \\
        $ T $ & N & Normalized time range \\
        $ F $ & $ 2\pi $ & Normalized frequency range \\
        $ D $ & $ \lfloor \frac{1}{2} \log_a N \rfloor $ & Number of levels of decomposition \\
        $ k $ & $ k \in [0, D - i_0), k \in \mathbb{Z} $ & Scale (time-spread) index \\
        $ m_k $ & $ 4a^{2k} $ & Number of chirplets at each scale \\
        $ M $ & $ N^2 \left( i_0 + \sum_k m_k \right) $ & Total number of chirplets in dictionary \\
        $ m $ & $ m \in [0, m_k - 1], m \in \mathbb{Z} $ & Chirp rate/rotational angle index \\
        $ \alpha_m $ & $ \arctan(m / a^{2k}) $ & Discretized angle for each scale \\
        \midrule
        $ t_c $ & $ t_c \in \gamma \frac{T}{N} = \gamma $ & Discrete time-center of chirplets \\
        $ \omega_c $ & $ \omega_c \in \gamma \frac{F}{N} = 2\pi\gamma / N $ & Discrete frequency-center of chirplets \\
        $ c $ & $ \frac{F}{T} \tan(\alpha_m) = \frac{F}{T}\frac{m}{\Delta_t} $ & Discrete chirp rate \\
        $ \Delta_t $ & $ a^{2k} $ & Discrete time-spread \\
        \bottomrule[1.5pt]
        
        \end{tabular}
    \end{center}
    \label{tb:cons_dic}
\end{table}
For a signal with size $ N $, the number of decomposition levels $ D $ is determined from $ N $ and the radix $ a $.  The first level in the decomposition is denoted as Level Zero.  Next, the scale index $ k $ and angle index $ m $ are calculated, from which the discrete chirp-rate $ c $ and time-spread $ \Delta_t $ are found.  The time-center $ t_c $ and frequency-center $ \omega_c $ are directly determined by the signal size $ N $.  The parameter $ i_0 $ indicates the first level to rotate a logon.  The reason for introducing $ i_0 $ may be understood in this way: a chirplet is close to the unitary Gabor logon if its time-spread is close to one.  If so, there is little significance to rotate it.  The parameter $ i_0 $ may be used to avoid unecessary rotation of chirplet close to unitary logon. 

At each iteration, a \textit{single} (new) chirplet $ g_{I_n}$ and coefficient $ a_{I_n} $ are decided from $ R^Pf(t) $.  This is termed ``coarse estimation". The results are further optimized using a \textit{Newton-Raphson} method to refine the match.  The refined results are then subtracted from the signal and the steps are repeated to estimate a new chirplet from the residue $ R^{P+1} f(t) $. We emphasize that the adaptive nature of the mechanism of the algorithm comes from the optimal selection of the basis functions for decomposition. The parameters of these functions are predefined in the dictionary, which differs from the approach of adaptive filtering where the parameters are varied on a sample by sample basis.

In the case of estimating \textit{multiple} chirplets, we have adopted the expectation-maximization (EM) algorithm  \autocite{Dempster1977} to further refine the estimates of the chirplets, which follows a framework for estimating superimposed signals using the EM algorithm \autocite{Feder1988}. More specifically, assuming the number of chiprlets $ P $ in the signal is known, the EM algorithm consists of an expectation step (E-step) and a maximization step (M-step). (1) In the \textbf{E-step}, the error between the signal under analysis $ f $ and the initial signal estimated is computed as
\begin{equation} \label{eq:em_error}
    e = f - \sum_{n = 1}^P a_{I_n} g_{I_n},
\end{equation}
and the complete data are formed as
\begin{equation}
    y_n = a_{I_n} g_{I_n} + \beta_n e, \quad n = 1, \ldots, P,
\end{equation}
where 
\begin{equation} \label{eq:beta_const}
    \sum_{n=1}^P  \beta_n = 1.
\end{equation}
(2) In the \textbf{M-step}, the same algorithm employed in the estimation of a single chirplet is applied to each of the $ y_n $ to refine the estimate of the corresponding chirplet. To alleviate computational cost, however, we follow the procedure proposed by O'Neill et al \autocite{Oneill2000} to update only one chirplet at each EM iteration by defining $ \beta_n $ for the \textit{i}-th iteration as $ \beta_n^{(i)} = \delta(n \text{ mod } i)$, where $ \delta(\cdot) $ designates the Kronecker function. The EM algorithm may be repeated several times until the change of error in \cref{eq:em_error} is below a threshold, or the number of iteration reaches a predefined level.   We'd like to point out that the $ \beta_n $'s may use other values, other than the one we suggested, as long as they satisfied the constraint stated in \cref{eq:beta_const}.  As is for all applications of EM algorithm, the selection of the complete data is crucial to the performance of specific algorithm.  Since the  $ \beta_n $'s determines the complete data in our algorithm,  they may influence the convergence of the algorithm and possibly be used to avoid the convergence to unwanted local stationary point.  Moreover, re-estimation in each $ y_n $ may be computed in parallel whenever the computer architecture of parallel computing is available.  These considerations are currently under investigation.  Finally, we summarize the MPEM algorithm in \cref{tb:mpem}. 
\begin{table}[t]
    \caption{The MPEM algorithm}
    \begin{center}
        \begin{tabular}{cl}
        
        \toprule[1.5pt]
        Step & Description \\
        \midrule
        1 & Construct chirplet dictionary (\cref{tb:cons_dic}) \\
        2 & Initialize residue: $ P \leftarrow 0 $, $ R^Pf \leftarrow f $ \\
        3 & Estimate a \textit{single} (new) chirplet \\
           & \hspace{6pt}3a. Estimate one chirplet from $ R^Pf $ with {\bf MP} algorithm \\
           & \hspace{6pt}3b. Refine with \textit{Newton-Raphson} (NR) method \\
           & \hspace{6pt}3c. $ P \leftarrow P+1 $ \\
        4 & Refine \textit{multiple} chirplets with EM algorithm \\
           & \hspace{6pt}4a. Initialize iteration counter: $ i \leftarrow 0 $ \\
           & \hspace{6pt}4b. {\bf E-step}: $ e \leftarrow f - \sum_{n=1}^P a_{I_n}g_{I_n} $; $ y_n \leftarrow  a_{I_n}g_{I_n} + \beta_n^{(i)}e $ \\
           & \hspace{6pt}4c. {\bf M-step}: Update $ a_{I_n} $ and $ g_{I_n} $ in $ y_n $ with MP+NR \\
           & \hspace{6pt}4d. $ i \leftarrow i+1 $ \\
           & \hspace{6pt}4e. Goto Step 4a, if stop criteria are not met. \\
        5 & Update $ R^Pf \leftarrow f-\sum_{n=1}^P a_{I_n}g_{I_n} $ \\
        6 & Stop MPEM if criteria are met; Otherwise goto Step 3. \\
        \bottomrule[1.5pt]
        
        \end{tabular}
    \end{center}
    \label{tb:mpem}
\end{table}

\begin{figure}[t]
\begin{center}
    \includegraphics[scale=0.70]{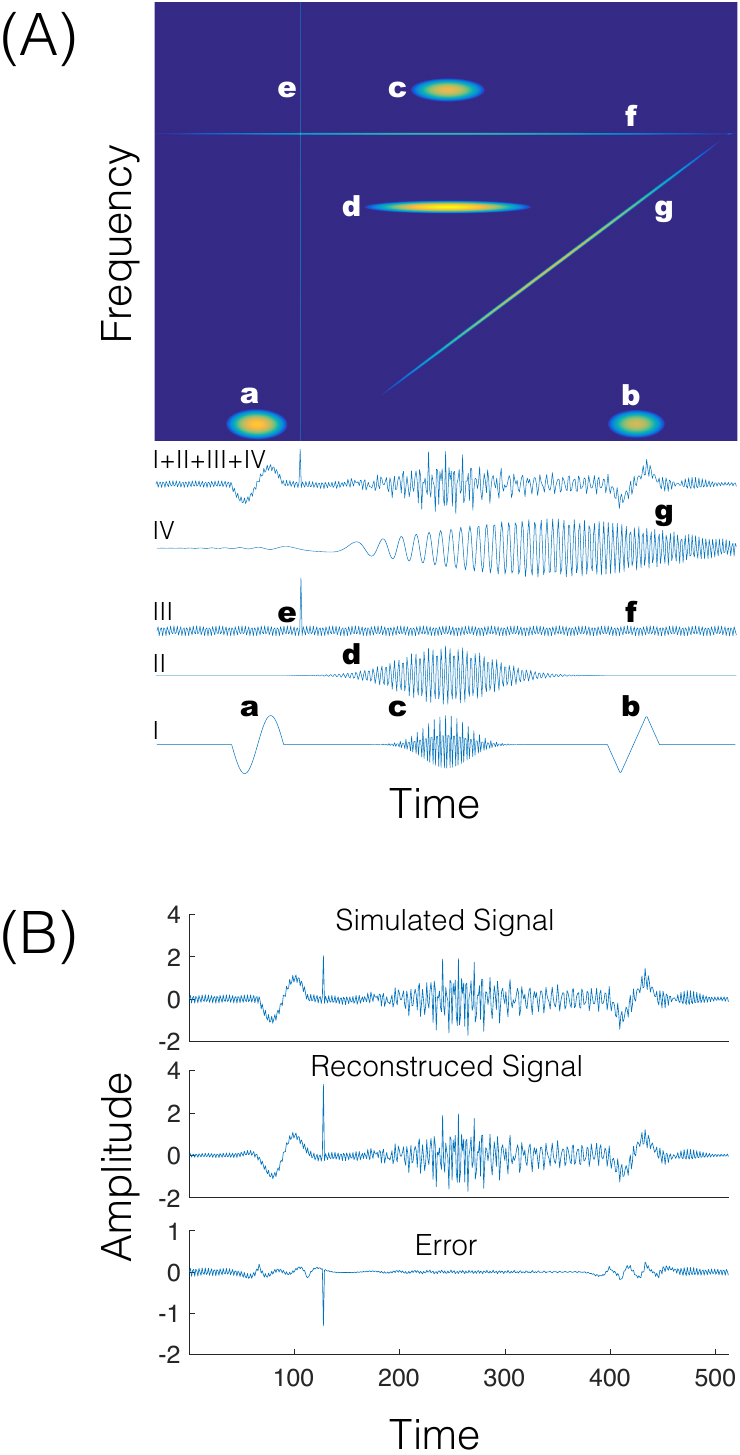}
\caption{\textbf{Decomposition of the simulated signal.} Panel \textbf{(A)} illustrates the simulated signal (waveforms at the lower portion of the panel) and the ACS of the estimated chirplets (the upper portion of the panels).  The synthetic signal  consists of seven (7) components, namely \textbf{a} = one period of a sinusoid,  \textbf{b} = one period of a saw-tooth wave, \textbf{c} and \textbf{d} = sinusoids modulated by a Gaussian, \textbf{e} = delta function, \textbf{f} = sinusoid and \textbf{g} = Gaussian chirplet (cf. the code provided online for the detailed description of the parameters). Panel \textbf{(B)} shows the original synthetic signal (top), the reconstructed signal obtained from the seven estimated chirplets (middle), and the error of the difference between the original signal and the reconstructed one (bottom).}
\label{fig:sim_decomp}
\end{center}
\end{figure}

The results shown in \cref{fig:sim_decomp} demonstrate the performance of the technique (cf. \textit{MultiDecompChirplet.m} in the code).  In this analysis, we show the decomposition of a complex signal into a number of Gaussian chirplets with MPEM algorithm.  The adaptive chirplet spectrogram (ACS) - which is a direct sum of the Wigner-Ville distribution of the individual chirplets - clearly shows the time-frequency structures of the signal.  Note the large error of representing delta function `\textbf{e}'.  This is because the delta function $ \delta(t) $ is not included in the dictionary, and thus the spike-like structure is approximated by a Gaussian chirplet with small time-spread.  The error of representing saw-tooth wave structure `\textbf{b}' is also larger than that of sinusoid structure `\textbf{a}', because Gaussian chirplets are able to approximate sinusoid better than saw-tooth function.  We'd like to emphasize that  the chirping structure `\textbf{g}' cannot be represented efficiently using Gabor logons alone (cf. \cite{Cui2006b}, p.41).  

In practice, a critical point of the analysis is to determine the number of chirplets required to sufficiently represent the signals.  One method is to employ the coherent coefficients ($ cc $) \autocite{Mallat1993} of the extracted chirplets, which is defined as the ratio of the energy of the projection to the energy of the residue.  The more coherent a signal is with respect to the dictionary, the larger the $ cc $ values are. Therefore, a small $ cc $ value indicates low correlation between the signal and the dictionary.  A threshold based upon the $ cc $ value can be chosen as a stopping criterion.
\begin{equation}
    cc_n = \frac{|a_{I_n}|^2}{|| R^nf ||^2}, \quad n = 0, \ldots, P-1,
\end{equation}
where $ |a_{I_n}|^2 $ is the energy of the projection and $ || R^nf ||^2 $ is the energy of the residue.  

\subsection{Robustness in low SNR}
As is discussed above, in general biological signals are collected under low SNR condition.  Thus the effectiveness of the proposed method to estimate signals in low SNR is an important factor in practice.  In this session, we quantify the robustness of MPEM algorithm under different levels of SNR by comparing it with the algorithm adopting the framework of maximum likelihood estimation (MLE) \autocite{Oneill1998}.  The MLE algorithm also adopts the ``coarse-refinement" scheme by first estimating the duration and the frequency-center of a chirplet, and then, from these local estimation, refining the estimation of the chirp-rate and the time-center.  We choose this algorithm to compare, since it has been widely cited in literature, has relatively low cost of computation, and possesses apparent robustness in high noise (as MLE algorithm avoids derivatives).

\begin{figure}[t]
\begin{center}
    \includegraphics[scale=0.42]{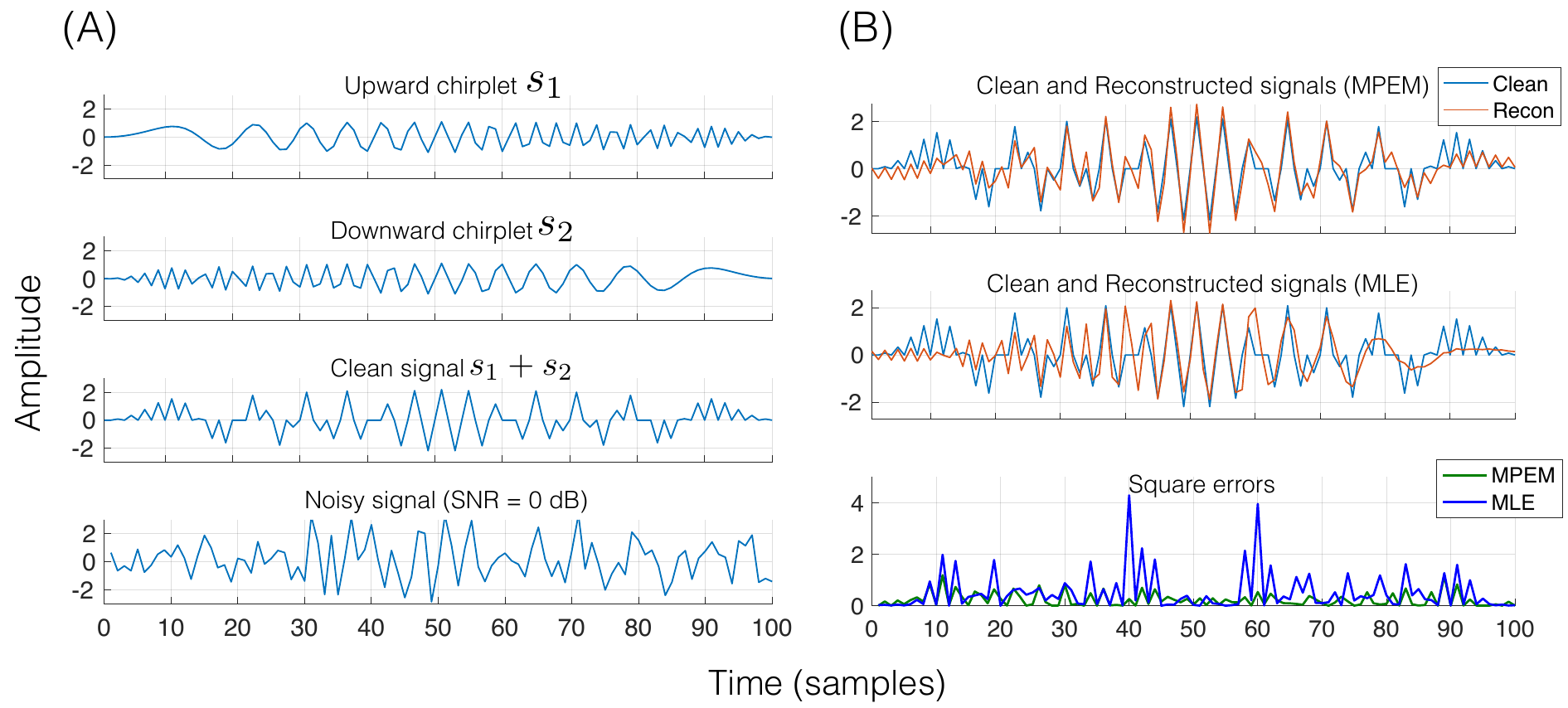}
\caption{\textbf{Simulation signal and an example of decomposition with MPEM and MLE algorithms.} Panel \textbf{(A)} shows the waveforms of the simulated signal, consisting of an upward and a downward chirplet, embedded in strong noise.  The top plot is the upward chirplet with the chirp-rate changing from zero to $ \pi $, and the second plot is the downward chiplet with the chip-rate changing from $ \pi $ to zero, and the signal length is 100 points. The third plot is the synthetic, clean signal, consisting of the two chirplets.  The bottom plot is an instance of the synthetic signal embedded in noise of SNR = 0 dB.  Panel \textbf{(B)} shows a typical example of the robustness of the algorithms MPEM and MLE against noise.  The top plot displays the original clean signal (``Clean") superimposed with the reconstructed signal (``Recon") estimated by MPEM algorithm from the noisy signal.  The middle plot is the same as the top plot except that the reconstructed signal is estimated by MLE algorithm.  The bottom plot compares the point-wise squared error produced by the two algorithms.  Note that MLE algorithm typically induces larger errors than MPEM algorithm at most of the time points.}
\label{fig:cross_chirplets}
\end{center}
\end{figure}

The relative robustness against noise is indicated by the Robustness Index (RI), a function of squared errors of the two algorithms,
\begin{equation} \label{eq:robust_idx}
R = \frac{E_l-E_p}{E_l+E_p},
\end{equation}
where $ E_l $ and $ E_p $ are the squared errors of MLE and MPEM algorithms, respectively. To calculate the squared error, we synthesized the simulation signal and embedded it in different levels of Gaussian white noise.  The simulation signal consists of an upward chirplet $ I_{u} = (N/2+1, \pi/2, \pi/N, N/3) $ and a downward chirplet $ I_{d} = (N/2+1, \pi/2, -\pi/N, N/3) $, where $ N $ is the signal length and the amplitude of both chirplets are set to one (\cref{fig:cross_chirplets}; cf. {\it noise\_robustness\_exp.m} in the code).
These two chirplet components share the same time and frequency centers, albeit have opposite chirp-rates.  The signals with such ``deep crossed" time-frequency structure is reportedly difficult for some coarse-refinement algorithms \autocite{Lyu2015}.  The chirplet components and their corresponding reconstructed signals have been estimated from the noisy signal using the two algorithms. The squared error is then found from the difference between the reconstructed and the clean synthesized signal per testing SNR level.  \cref{fig:cross_chirplets} presents a typical example of multiple chirplet decomposition of a noisy signal with MPEM and MLE algorithms.  We can see that MLE incurs larger errors than MPEM algorithm at most of the time points.  

\begin{figure}[h]
\begin{center}
    \includegraphics[scale=0.55]{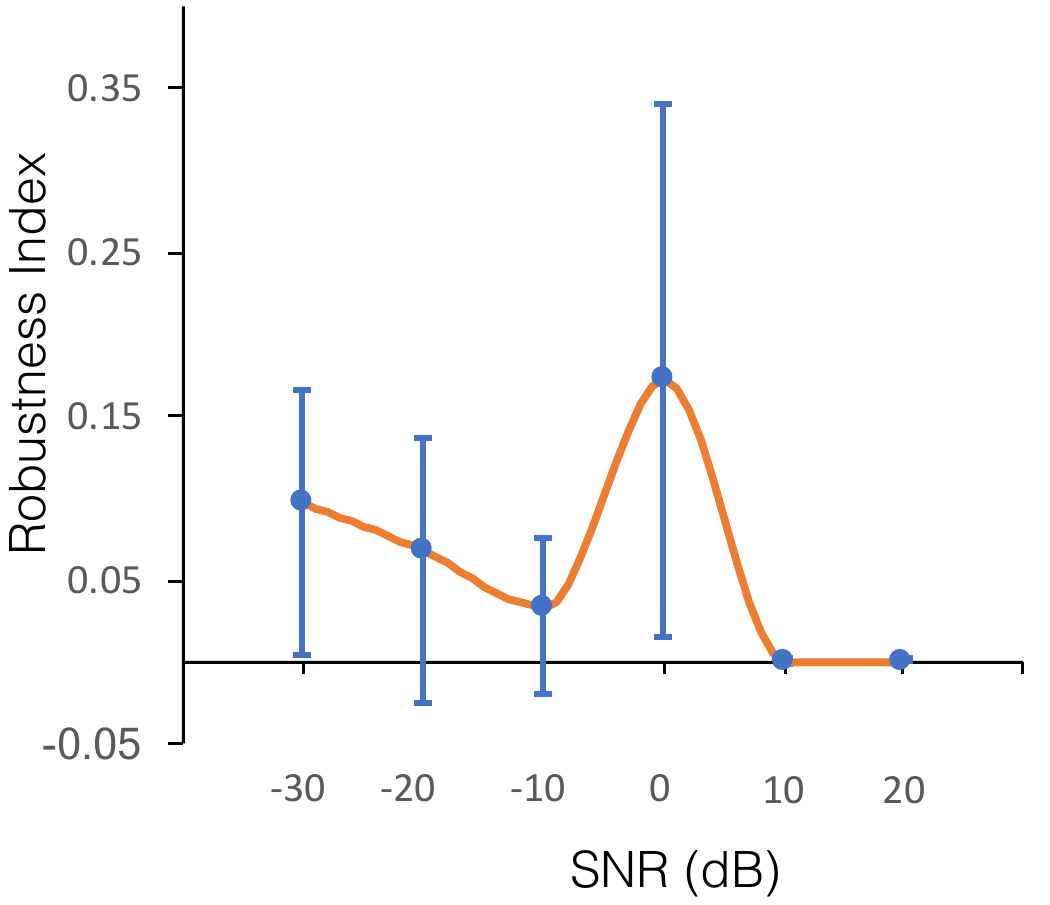}
\caption{\textbf{Testing the robustness of the algorithms against different levels of noise using the Robustness Index (RI).}  The relative robustness of the two algorithms against noise of different levels is measured as RI (\textit{ordinate}), \cref{eq:robust_idx}, at each testing SNR point (\textit{abscissa}).  The error bar indicates the 95\% confidence interval (the intervals are smaller than the dot sizes at 10 dB and 20 dB).  The higher the index, the higher the robustness of MPEM algorithm than MLE algorithm is.  The RIs between the testing SNR points are indicated as the smooth curve obtained with a shape-preserving piecewise interpolation method.  Notice that statistically at all test points the robustness of MPEM algorithm is significantly higher than MLE algorithm (see the text).}
\label{fig:rb_index}
\end{center}
\end{figure}

Notice that RI is between -1 and 1, i.e. $ |I| \leq 1$, \cref{eq:robust_idx}.  If the error of MLE algorithm is higher than that of MPEM algorithm (i.e. $ E_l > E_p $), RI will be greater than zero. Thus, the higher RI is, the larger the squared error of MLE algorithm ($ E_l $) than that of MPEM algorithm ($ E_p $) is, which means the robustness of MPEM is better than MLE.  We formulated the statistics of RI by using Bootstrap resampling technique \autocite{Efron1993} as follows.  First, the noisy signal was created by adding additive Gaussian white noise at five SNR levels, i.e. -30, -20, -10, 0, 10 and 20 dB, as biosignals are usually collected within these SNR ranges.  Second, at each testing SNR, 100 RIs were calculated.  Finally, the mean and 95\% confidence interval were estimated by resampling the RIs for 1000 times with the Bootstrap approach.  The results are presented in \cref{fig:rb_index} (cf. {\it noise\_robustness\_analysis.m} in the code).  
The scatter plot of dots indicate the mean values of RIs, and the error bars the confidence intervals.   The RIs between the testing SNR levels are indicated with a smoothed curve approximated by shape-preserving piecewise cubic interpolation, assuming a smooth change of RIs as a function of SNR. We can see that, with the decrease of SNR, RIs increase significantly above zero, which indicate that the robustness of MPEM is higher than that of MLE in stronger noise.  However, the change of RI is not monotonic, but with a peak around SNR = 0 dB.  That is, the relative increase of performance of MPEM algorithm is strongest when the energy of signal and noise is roughly equal.  At higher SNR levels, i.e. 10 and 20 dB, the RIs are close to zeros, indicating that when noise is low, the performance of the two algorithms are approximately equal.  We also performed Wilcoxon Rank Sum test with the null hypothesis that RI is not significantly higher than zero (Right-tailed hypothesis test).  We found that the null hypothesis was rejected at all testing SNR levels ($ p < 10^{-4} $), which indicates that the robustness of MPEM algorithm is consistently higher than that of MLE algorithm.  However, we should point out that the superior performance of MPEM against noise is not achieved without consequence.  The average time cost of MPEM algorithm is significantly higher than MLE algorithm.  To extract the two chirplets of the simulation signal, the time MPEM requires is $ \sim5.46 $ folds higher than MLE method (tested on iMac-Late 2014 equipped with 3.5 GHz Intel Core i5 processor, 24 GB 1600 MHz DDR3 memory and AMD Radeon R9 M290X graphics).

\section{Applications to biosignal analysis}
In this section we demonstrate the capability of our method by applying it to a set of selected biosignals.  Most biological systems are accompanied by or manifested themselves as \emph{signals} that reflect the nature of their normal or abnormal processes.  However, biosignals are notorious for their variability, or non-stationarity, and thus joint time-frequency analysis plays a major role in biosignal processing.  Given the abundance of chirping phenomena in biological systems, the adaptive chirplet transform (ACT), as a newly emerged tool of time-frequency analysis, has potential applications in the analysis of signals that involve frequency-changing components.  Here we show the efficiency of the MPEM algorithm in the analyses of some representative data from two broad categories of biosignals: bioelectrical signals (e.g. visual evoked potentials) and bioacoustical signals (e.g. heart sounds, bat echolocation signals, bird songs and human speech).  

\subsection{Apply ACT to visual evoked potentials}
Visual evoked potentials (VEPs) are scalp electrical signals generated in response to rapid and repetitive visual stimuli.  VEPs have prominent clinical significance and can help diagnose sensory dysfunctions. They are traditionally employed in testing the integrity
of the visual pathway and used as a supplement to other techniques in research into specific clinical conditions.  A variety of clinical applications require the analysis of the steady-state visual evoked potentials (\textit{ss}VEPs), which are elicited when the the repetition rate of visual stimuli is sufficiently high \autocite{Regan1989}.  We describe an application of the ACT to the analysis of \textit{ss}VEP signals.  The goal of these approaches is to characterize the time-dependent behavior of VEP from its initial transient portion to the steady-state portion by a series of time-frequency atoms, or chirplet basis functions. The ACT technique allows us to clearly visualize, perhaps for the first time, the early moments of a VEP response.

The VEPs were obtained through experiments involving a matrix of moving bars aligned to visual fixation crosses. The details are described elsewhere \autocite{Cui2006a}.  Briefly, an averaged signal was obtained from 50 trials of a single subject.  All data were collected in accordance with the protocol of human experimentation of the University of Toronto, Ontario, Canada. Ten Gaussian chirplets were estimated with the algorithm described in \cref{sec:gct}.  We believe that this number of chirplets is sufficient for representing the VEP of interest as the residue after ten iterations was virtually indistinguishable from white noise.  In general, the chirplets extracted first
have higher amplitudes and higher $ cc $. Particularly, we found that the amplitudes of the first three chirplets are significantly higher than those of the remaining chirplets.  
\begin{figure}[t]
\begin{center}
    \includegraphics[scale=0.50]{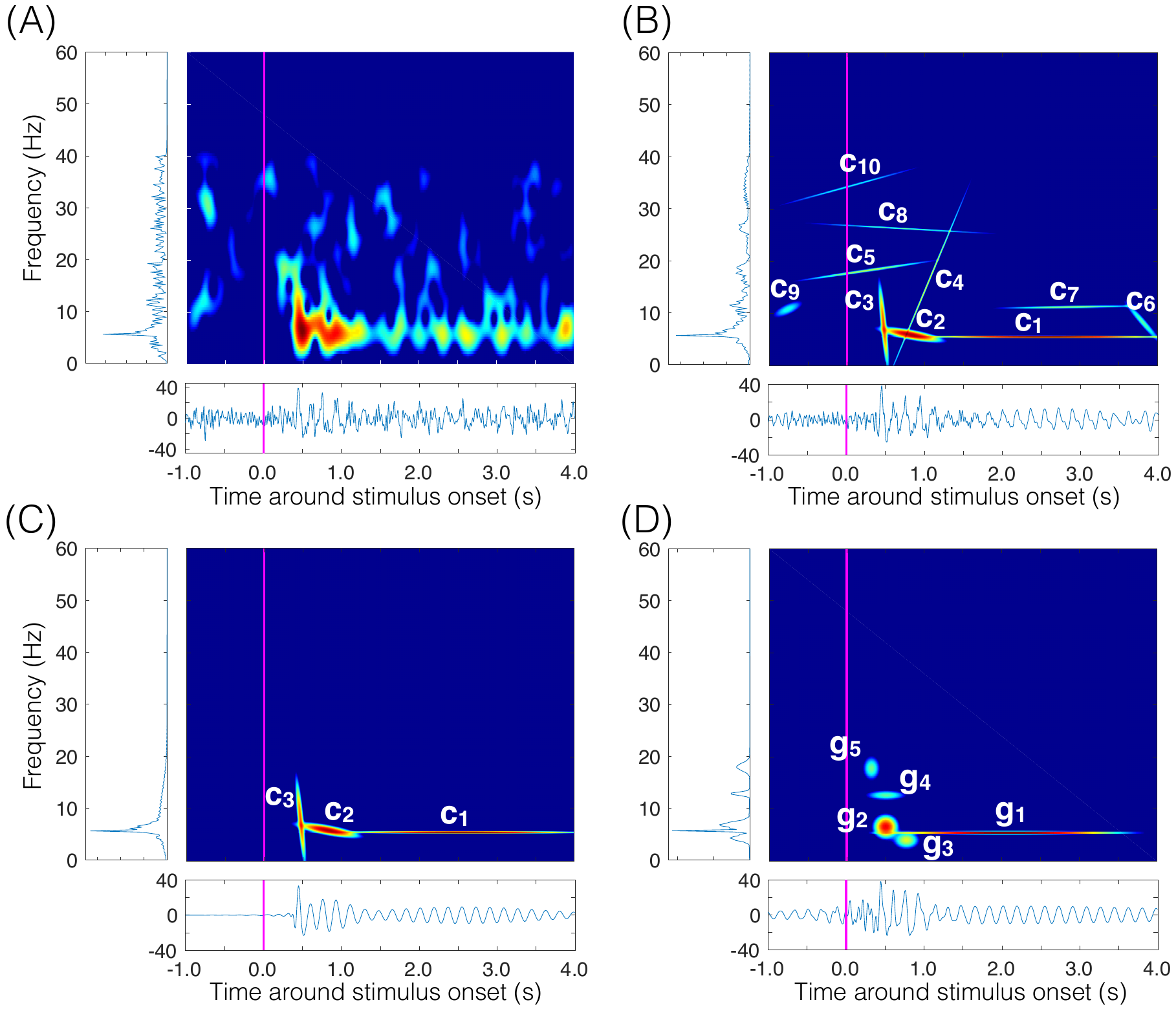}
\caption{\textbf{Time-frequency analysis of visual evoked potentials (VEPs).} Panel \textbf{(A)} shows the spectrogram (short-time Fourier transform with 11-point Gaussian window) of the VEPs.  The waveform of the original VEP signal is presented immediately below the time-frequency representation, while the corresponding spectrum of the signal is presented on the left side.  Panel \textbf{(B)} shows the ACS of the 10 estimated \emph{Gaussian chirplets} of the VEPs by using the MP based adaptive chirplet transform (MPEM algorithm).  The reconstructed signal from these 10 chirplets is shown below and the spectrum of the reconstructed signal on the left.  Note  the transient phase of the response evoked immediately after the stimulus onset, which possesses the decrease of instantaneous frequency from high to low in less than one second, and the steady-state phase in the later portion of the response.  This characteristic transition from initial to steady response can be better represented concisely by as few as three chirplets, as is shown in Panel \textbf{(C)}.  The three chirplets have the highest correlation coefficients among the estimated ones.  As a comparison, Panel \textbf{(D)} displays the time-frequency distribution of \emph{Gabor logons} of the VEPs.  Notice that at least five logons are needed to adequately characterize the transition and steady-state phase of the evoked potentials. The pink vertical line indicates the onset of visual stimulus.}
\label{fig:act_vep}
\end{center}
\end{figure}
As shown in \cref{fig:act_vep}, the first chirplet, $ c_1 $, represents the steady-state component of the VEP signal, as it has a long time-spread ($ \Delta_t $) and near zero chirp rate. The remaining two chirplets $ c_2 $ and $ c_3 $ have negative chirp rates, indicating that the instantaneous frequency of the prominent early components decreased with time.  \cref{fig:act_vep} also shows the visualization of the results using ACS and compares it to the conventional spectrogram calculated with STFT.  Panel (B) shows the ACS of the ten chirplets accompanied by the reconstructed signal shown directly below and the spectrum on the left.  It can be seen that the reconstructed signal provides a \emph{less noisy} waveform.  Chirplet $ c_1 - c_3 $ are shown separately in Panel (C), a typical representation of {\it ss}VEPs.  In Panel (D) the signal was approximated with the traditional Gabor logons.  The transient portion was represented by four Gabor logons ($ g_2 - g_5 $), instead of two chirplets ($ c_2 $ and $ c_3 $) as shown in Panel (C), which demonstrates the efficiency of chirplet representation.  

By adopting the ACT approach, we can thus achieve a very sparse representation of the original VEP signal.  Furthermore, the conventional information of EP analysis, such as amplitude and latency, of the signal has been retained and can be retrieved readily from the reconstructed signal.  The ACS allows us to visualize the time-frequency structure of the VEP response at higher resolution than previously possible.  Spectrograms that have been constructed using STFT will invariably involve smoothing of some sort yielding an overall lower resolution picture.  Although the spectrogram can show some of the salient time-frequency structures of the VEP response, most of the detail is lost due to smearing. However, as we have shown with the ACS, e.g., (C) in \cref{fig:act_vep}, the resulting time-frequency decomposition provides a clear picture of the underlying process. The estimated parameters obtained from the decomposition analysis provide detailed information about the local time-frequency structures of the signal, which are not easily obtainable from the standard spectrogram alone.

\subsection{Apply ACT to heart sounds}
The phonocardiogram (PCG), or \emph{heart sounds}, is perhaps the most traditional biomedical signal, as indicated by the fact that the stethoscope is the primary instrument carried by physicians.  The normal heart sounds provide an indication of the general state of the heart, while cardiovascular disease and defects cause changes or additional sounds and murmurs that could be useful in diagnosis.  In a normal cardiac cycle one may hear two distinctive sounds --- the first (S1) and second (S2) heart sound.  The epoch of S1 is directly related to the event of ventricular contraction, reflecting a sequence of events related to closure of cardiac valves and ejection of the blood from the ventricles.  The epoch of S2 reflects a series of events related to the end of ventricular contraction, signified by closure of the aortic and pulmonary valves.

\begin{figure}[t]
\begin{center}
    \includegraphics[scale=0.55]{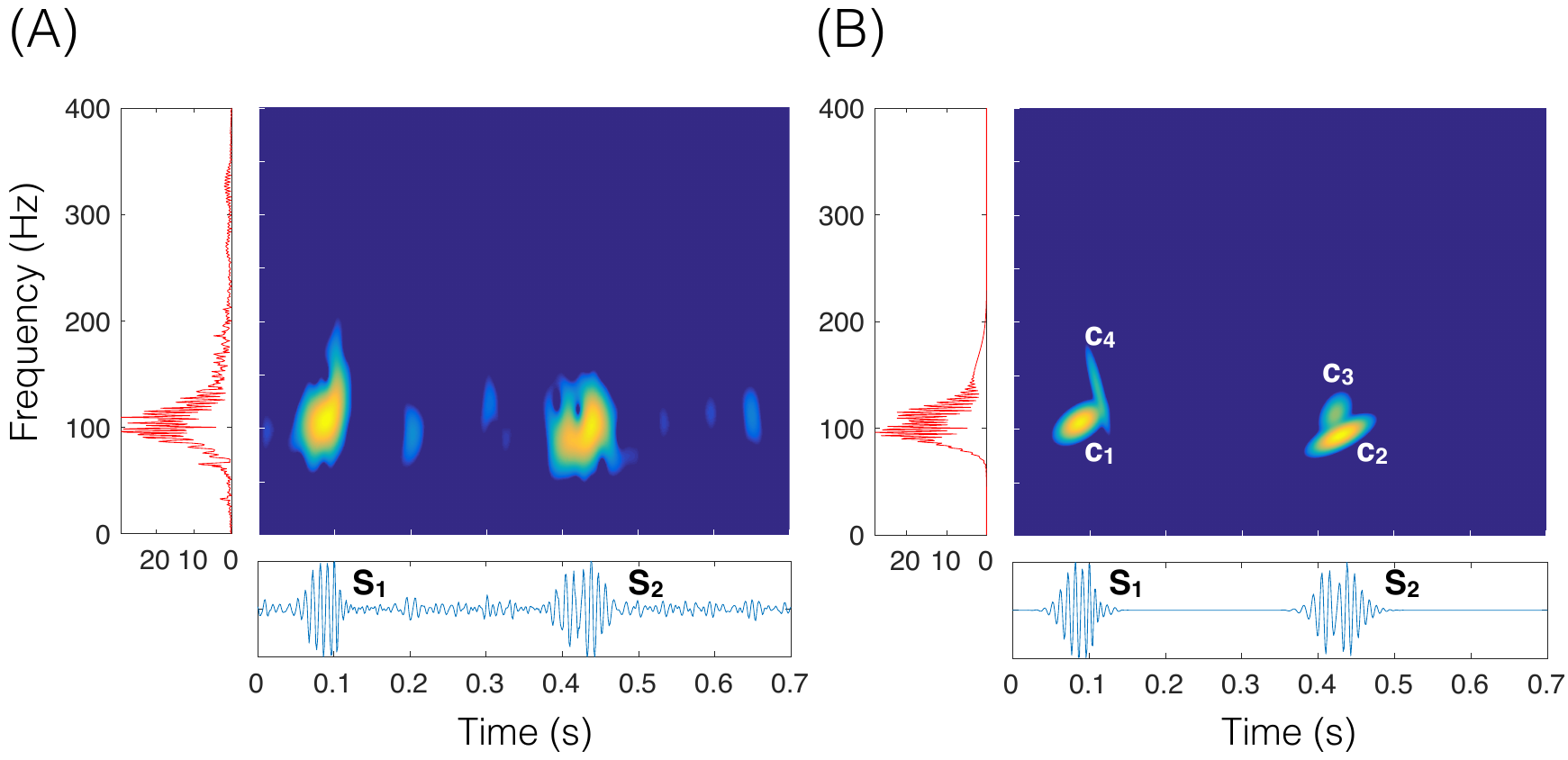}
\caption{ \textbf{Time-frequency analysis of heart sounds.} Panel \textbf{(A)} shows the spectrogram of a recorded heart sounds within one cardiac cycle.  The first heart ($ \mathbf{S_1} $) and second heart sound ($ \mathbf{S_2} $) can be clearly identified.  A four-chirplet representation of the heart sound is shown in Panel \textbf{(B)}, where each major component  is represented by two Gaussian chirplets ($ \mathbf{c_1} $, $ \mathbf{c_4} $ for $ \mathbf{S_1} $, and $ \mathbf{c_2} $, $ \mathbf{c_3} $ for $ \mathbf{S_2} $).  Importantly, since the chirp-rates of these chirplets significantly deviates from zero, the chirplet analysis indicates the frequency-changing character of the sound components.}
\label{fig:act_heart}
\end{center}
\end{figure}

The frequency contents of heart sound have a long history in diagnosis, which are believed in significant value in the evaluation of the heart condition.  As heart sounds are non-stationary in nature, more recent research adopts the time-frequency analysis in order to capture the temporal variation in the heart sound signals.  The analysis of heart sounds by MP approach has been proposed before \autocite{Zhang1998}.  However, the dictionary employed in the previous study was composed of Gabor logons, and thus the frequency changing components could not be represented efficiently.  Here, we demonstrate the capability of ACT in phonocardiogram analysis. The signal of heart sounds\footnote{The original signal is available through Department of Medicine at University of Washington(https://depts.washington.edu/physdx/heart/demo.html)} consists of one cycle of heart beat with a duration of 700 ms.  The original signal was sampled at 8 kHz.  Since most of the diagnosis relevant components of heart sounds are below 300 Hz, however, we further downsampled the signal to 800 Hz, resulting in a signal of 700 sample points.  The results are shown in \cref{fig:act_heart}.  
Panel (A) shows the spectrogram (estimated by STFT) of a heart sound signal. The waveform of the signal in the time domain is immediately shown below and its spectrum in the frequency domain is shown on the left.  The first (S1) and second (S2) heart sound are represented by two distinctive energy blobs in the spectrogram, but their detailed time-frequency structures are not readily appreciated.  On the other hand, Panel (B) displays the ACS of the four major chirplets extracted from the signal.  We can see that the dominant components, i.e., \textbf{c\textsubscript{1}} of S1 and \textbf{c\textsubscript{2}} of S2, show the positive chirp-rates, indicating an increment of frequency in the beat.  This information was not clearly available with neither the conventional spectrogram,  nor the previous approach using Gabor logons \autocite{Zhang1998}.  

\subsection{Apply ACT to bio-acoustical signals}
\begin{figure}[t]
\begin{center}
    \includegraphics[scale=0.60]{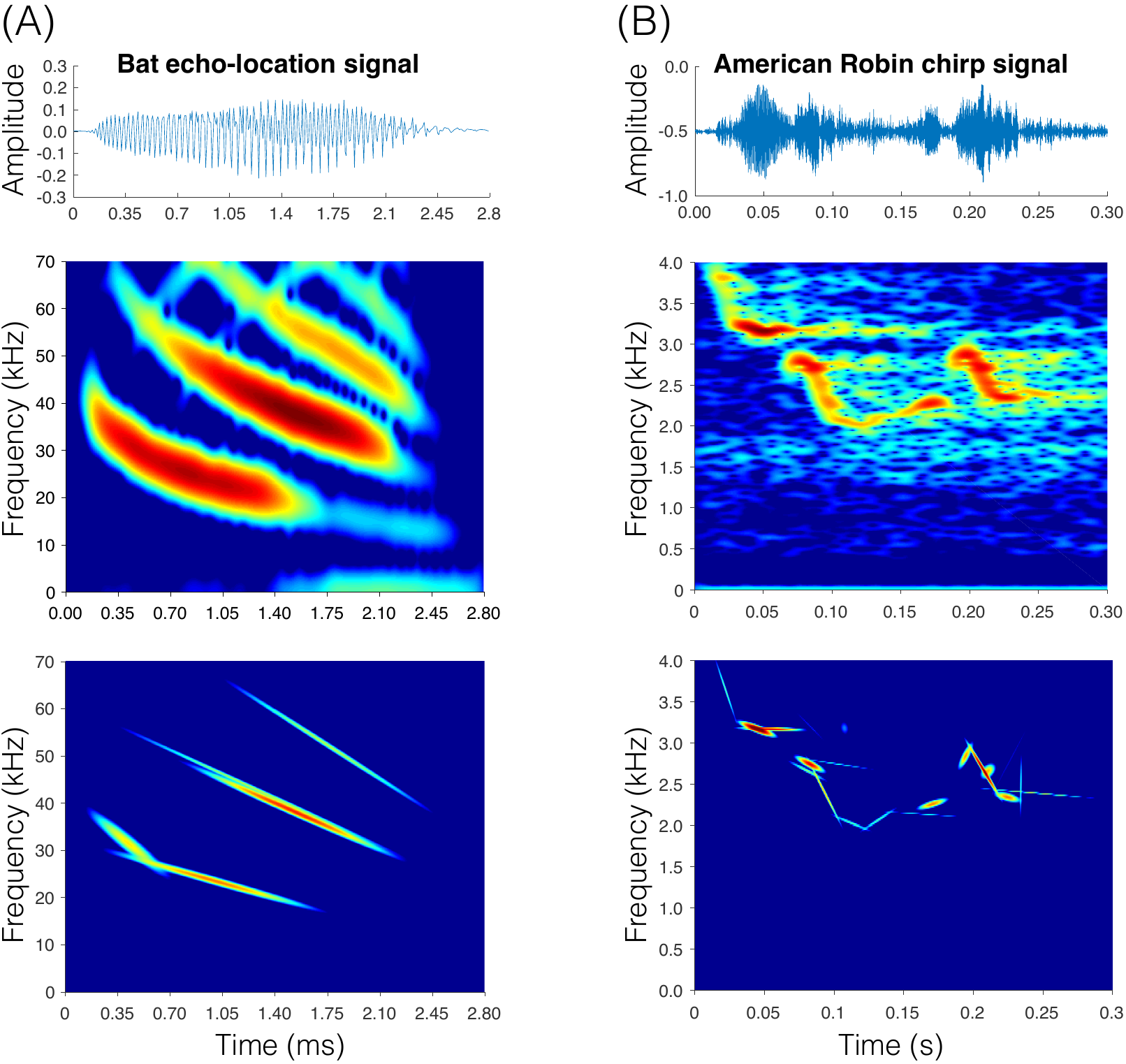}
\caption{\textbf{Chirplet representation of bio-acoustical signals.}  Panel \textbf{(A)} demonstrates the application of MPEM ACT to the analysis of an ultrasonic bio-signal.  The top plot shows the time domain waveform of an echo-location signal of large brown bat (sampling frequency $ \approx $\SI{140}{\kilo\hertz}).  The middle image is the spectrum of the signal (calculated with a 0.45 ms Gaussian window), and bottom image the adaptive chirplet spectrum (ACS) representation (consisting of five chirplets).  Panel \textbf{(B)} shows another demonstration by analyzing an audiable sound - the chirping sounds of American Robin.  The top plot is the waveform of the bird song in the time domain (sampled at 8 kHz), the middle one the spectrogram (with an 8 ms Gaussian window), and the bottom one the ACS represented by 20 chirplets. }
\label{fig:act_bioacoustics}
\end{center}
\end{figure}

As mentioned in \cref{sec:intro} Introdcution, bio-acoustical signals are full of chirping components.  Two examples of chirplet representations are shown in \cref{fig:act_bioacoustics}.  
The signals are a bat echo signal and a bird whistle of American Robin\footnote{At the courtesy of C. Condon, K. White, and A. Feng of the Beckman Center at the University of Illinois}, which are in the ultrasonic range and audible frequency range, respectively.  The bat signal was sampled at \SI[mode=text]{7}{\micro\second} for a duration of \SI{2.8}{\milli\second}, and the bird song was sampled at \SI{8}{\kilo\hertz} for a duration of \SI{3}{\second}.  Panel (A) shows the results of the analysis of bat echo-location signal. The top panel shows the waveform of the bat signal in the time domain, and the middle is the corresponding spectrogram.  The bottom illustrates the ACS of the five estimated chirplet atoms with the proposed MPEM algorithm.  We can see that the chirplet representation can clearly show the major time-frequency structures of the ultrasonic signal.  Panel (B) demonstrates the capability of the ACT in the audible frequency range.  The chirp signal (top plot) of an American Robin was represented by 18 chirplets.  Not only can the ACS (bottom plot) provide a \emph{clearer} visualization of the energy content of the bird song, but also the reconstructed signal produces a \emph{cleaner} sound perceptually.  These examples demonstrate the value of the compact representations using chirplets.

\subsection{Apply ACT to speech signal}
As a novel tool of time-frequency analysis, the ACT has potential applications in the analysis of human speech. 
\begin{figure}[t]
\begin{center}
    \includegraphics[scale=0.58]{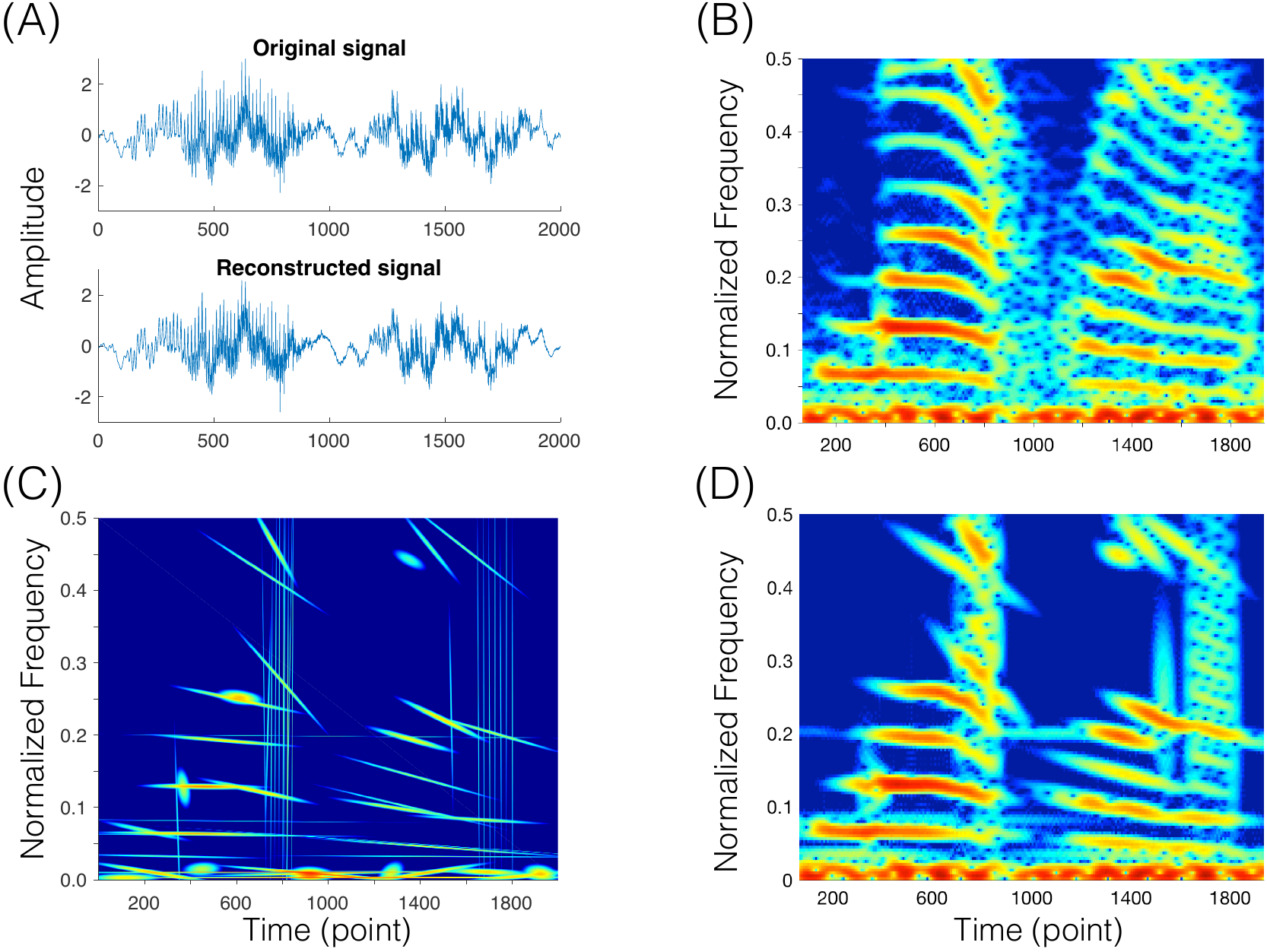}
\caption{ \textbf{Chirplet representation and compression of speech signal.} Panel \textbf{(A)} displays the waveform of an acoustic speech signal of the spoken word "Matlab",  of which the top plot shows the original signal and bottom one the reconstructed signal from the 60 estimated chirplet components. The spectrogram (STFT) of the original signal is shown in Panel \textbf{(B)} and the ACS of the speech signal represented by 60 chirplets is shown in Panel \textbf{(C)}.  As a comparison, Panel \textbf{(D)} shows the spectrogram of the reconstructed signal.}
\label{fig:act_speech}
\end{center}
\end{figure}
\cref{fig:act_speech} shows a chirplet analysis of a female utterance of the word ``Matlab"\footnote{The sound file was included in MATLAB\textsuperscript{\textregistered} 7 R14.}.  The signal length is 2000 points sampled at \SI{3.71}{\kilo\hertz}.  60 chiprlets were extracted and a reconstructed speech signal was obtained from these chirplets.  There is minimal perceptual difference between the original and the reconstructed signals.  The waveforms of the original and the reconstructed signals are shown in Panel (A) of \cref{fig:act_speech}.  Panel (B) displays the STFT of the original signal.  Panel (C) shows the ACS of the extracted chirplet components, of which the STFT of the reconstructed signal is shown in Panel (D).  Recall that one chirplet can be described by six real-valued parameters, i.e., two for the complex coefficient in \cref{eq:gct} and four for chirplet parameters  $ I = (t_c, \omega_c, c, \Delta_t) $ defined in \cref{eq:gaussain_chirplet}.  Therefore, the 60 chirplets required only 360 real values for the entire data recording, showing an approximately 80\% reduction in data size when compared with the 2000 real values of the original signal.

\section{Discussion}
Biosignals are non-stationary in nature.  The phenomenon of varying frequency exists abundantly in biological systems.  The ACT, which in essence approximates the energy of a signal on the time-frequency plane by straight lines, is considerably promising for compactly representing biosignals.  We'd like to emphasize that despite the fact that the chirplet transform has been proposed for more than 25 years and that ACT is well accepted in the community of signal processing \autocite{Mann1992, Baraniuk1993}, the application of ACT to the field of biomedical engineering is still relatively limited.  A search with Google Scholar, by far, only yields a double-digit number of studies related to the application of chirplet analysis to biosignals.  One possible reason is that the researchers in the field have not fully appreciated the merits of ACT for the analysis and representation of non-stationary signals with biological nature.  In this article, we have introduced a new approach to the ACT, namely MPEM algorithm, in which the coarse estimation of the chirplet parameters is obtained with the \emph{chirplet} MP algorithm and subsequently, the parameters are refined with EM algorithm.  We have demonstrated the capacity of our algorithm by applying it to four representative, highly non-stationary biosignals, i.e. visual evoked potentials, heart sounds, bio-acoustic signals (ultrasonic bat echolocation signal and audible bird chirps), and human speech.  These results demonstrate the value of sparse representation with chirplet basis.  Signal information is diluted less and packed into a few coefficients of high energy, which makes ACT an appealing alternative to data compression for longterm signals (e.g., monitoring signals generated by the devices in ICU or ambulance).  Moreover, The ACS provides a clear picture of a signal's energy content, and thus captures the ``signature" of the signal in the time-frequency plane, which should be especially useful in pattern recognition problems, such as computer aided diagnosis.

One unique feature of our method of ACT is that, unlike other chirplet based MP algorithms that employ the dictionary of Gabor logons for their coarse estimation, our algorithm uses chirplet dictionary directly.  This approach avoids the difficulty of decomposition of multiple chirplets when their time-frequency centers cluster closely (so called \emph{deep crossed}  situation), which is a typical defect for the \textit{Gabor-to-chirplet} approach (see \cref{sec:intro}. Also, cf. \textit{DecompDeepCrossChirplet.m} in the code). Furthermore, the adoption of chirplet dictionary for MP algorithm leads to increment of robustness against noise, which is relevant to biomedical application, as biosignals usually have low SNR.  This advantage may be understood from a point of view of \emph{matched filters}.  It is known that the best detection of signals in the presence of noise is based on the matched filter output, which is essentially an inner product between the noisy measurements and the target signal \autocite{vantree2001}.  From the discussion in \cref{sec:gct} we know that the Gaussian chirplet transform (GCT) defined in \cref{eq:gct} is the inner product between the signal and elements of the dictionary.  Thus, we can think of the elements in the chirplet dictionary as many \emph{templets} used in the procedure of matched filtering.  Since Gabor logons is a subset of chirplets, the chirplet based MP has more templets to \emph{match} the time-frequency structure of the noisy signal and thus is more robust against noise.  

An important issue in future research on MP approach in general, and on chirplet based MP approach in specific, for biomedical application, is the \emph{construction} of the dictionary.  Previous studies have indicated that the choice of elements, or \emph{templates}, to form an overcomplete set to cover signal space is crucial to the possession of greater robustness in face of noise, and to the production of efficient coding of the biosignals.  For example, in an EEG study the generally structured, signal independent dictionary, could lead to biased representation (i.e. artifacts) of sleep spindles \autocite{Durka2001}. Interestingly, humans and animals may adopt some similar mechanisms to detect and perceive signals by matching them with biological ``templets" \autocite{Wong2000}, which, acquired through experience, are behaviorally relevant and environmentally adapted.  An efficient code is intrinsically entangled with the class of signals being encoded \autocite{Field1987}. Existing evidences suggest that somatosensory \autocite{Macfarlane2009}, visual \autocite{Olshausen1996} and auditory \autocite{Lewicki2002} cortexes may employ overcomplete basis sets to sparsely code \emph{natural} signals impinging on our sensory systems. Inspired by the scientific discoveries, recent work in engineering has been exploring the ways to find specific dictionary for a given set of signals \autocite{Aharon2006a, Aharon2006b, Donho2006}, which is aimed at seeking unique and stable representation of the signal of specific class in the presence of noise.  Particularly, the techniques of artificial intelligence and machine learning have been increasingly showing promise in this quest \autocite{Aharon2006a, Qiu2010}.  These techniques are expected to refine the dictionary relevant to the class of signals under investigation, and thus enhance the efficiency of sparse representation \autocite{Bengio2013}.

\section{Summary}
In this article, we have introduced a new approach to the adaptive chirplet transform (ACT) and emphasized its merits in the analysis of signals produced in biological systems.  Biosignals are non-stationary in nature and usually measured with low signal-to-noise ratio (SNR).  Our method employ a coarse-refinement strategy to alleviate the high cost of computation in chirplet transform.  Particularly, we adopt the \emph{chirplet} dictionary directly in the coarse step of estimation and the expectation-maximization (EM) algorithm to refine the parameters.  The matching-pursuit (MP) algorithm has been used to implement the multi-component extraction of chirplet atoms.  We have demonstrated the capability of our approach by applying it to some representative biosignals, including bioelectrical signals (i.e. visual evoked potentials), bio-acoustical signals (i.e. heart sounds, echolocation sounds) and human speech.  Our techniques result in more compact representation of these signals, clearer visualization of their time-frequency structures, and increase of estimation robustness in the face of strong noise, which shows considerable promise of chirplet analysis for biosignal processing.  Finally, we point out in the discussion that the technology developed in the field of machine learning would be crucial in the future to construct ``dictionary", or ``code book", for efficient coding of the biosignals under investigation.

\printbibliography


\end{document}